\documentclass[12pt, preprint]{aastex}
\newcommand{\eri}{$\epsilon$~Eridani}

\begin{document}
\slugcomment{Accepted by ApJ, May 13th, 2004}
\title{The Dynamical Influence of a Planet at Semimajor Axis
3.4 AU on the Dust Around \eri}
\shorttitle{\eri\ Dust Dynamics}
\author{Sean M. Moran}
\email{smm@astro.caltech.edu}
\affil{Caltech Astronomy Department, MC 105-24, Pasadena, CA 91125}
\author{Marc J. Kuchner\altaffilmark{1}}
\email{mkuchner@astro.princeton.edu}
\affil{Department of Astrophysical Sciences, Princeton University, Peyton Hall, Princeton, NJ 08544}
\author{Matthew J. Holman}
\email{mholman@cfa.harvard.edu}
\affil{Harvard-Smithsonian Center for Astrophysics,
60 Garden Street, Cambridge, MA 02138}
\altaffiltext{1}{Hubble Fellow}

\begin{abstract}
Precise-Doppler experiments suggest that a massive ($m \sin i=0.86 M_J$)
planet orbits at semimajor axis $a=3.4$~AU around \eri, a nearby star
with a massive debris disk.  The dynamical perturbations from
such a planet would mold the distribution of dust around this star.  We
numerically integrated the orbits of dust grains in this system to
predict the central dust cloud structure.  For a supply of
grains that begin in low-inclination, low-eccentricity orbits at 15~AU,
the primary feature of the dust distribution is a pair of dense clumps,
containing dust particles trapped in mean-motion
resonances of the form $n:1$.  These clumps appear to revolve around the star
once every two planet revolutions.  Future observations with the
IRAM Plateau de Bure Interferometer, the SMA, or ALMA could detect
these clumps, confirming the existence of the planet and
revealing its location.

\end{abstract}

\keywords{celestial mechanics --- circumstellar matter --- interplanetary medium --- planetary systems --- stars: individual (Epsilon Eridani)}

\section{INTRODUCTION}

Some nearby main sequence stars appear to host debris belts like
the asteroid belt and the Kuiper Belt in our solar system; see the
reviews by \citet{backman} and \citet{zuck01}. Extrasolar
asteroids and Kuiper Belt objects can reveal themselves by
generating clouds of circumstellar dust that emit thermally in
excess of the stellar photospheric emission. The Infrared
Astronomical Satellite (IRAS) discovered dozens of ``Vega-like''
stars which show signs of circumstellar dust, and upcoming
observatories like the Keck Interferometer, the Spitzer Space
Telescope, JWST and Darwin/TPF should detect many more.

The small body belts in our solar system bear
the dynamical signatures of planets; perhaps their extrasolar
analogs do too.  Millimeter and submillimeter images of dust
rings around the Vega-like stars $\beta$ Pictoris, Vega, Fomalhaut, and \eri\
all show clumps and asymmetries \citep{holland, greaves, koer01, wilner, holl03}.
Dust spiraling past a planet under the influence of Poynting-Robertson
(P-R) drag \citep{robertson,wyat50} can become temporarily trapped in
exterior mean-motion resonances (MMRs) with the planet, forming rings
of enhanced dust density \citep{gold75} like the one created by the
Earth in the solar zodiacal cloud \citep{dermott,reac95}.  Locating
clumps and holes in these rings can constrain the position, mass and
orbital eccentricity of the perturbing planet \citep{khsignatures,kuch03}.

Of the four Vega-like stars mentioned above, the K2 V star \eri\ is closest
to Earth at a distance of 3.22 pc.
\citet{greaves} imaged a blobby ring of emission around \eri\
at 850~$\mu$m using the  Submillimeter Common-User Bolometer Array
(SCUBA) on the James Clerk Maxwell Telescope (JCMT).  \citet{quillen}
have modeled the blobs in this ring, and suggest that a perturbing
planet at a semimajor axis of $\sim 40$~AU could be responsible for the
observed asymmetries.
One might consider this ring at $\sim 60$~AU to be an analog of the
Kuiper belt, perturbed by Neptune \citep{liou, moro03}---though
the \eri\ Neptune analog has an eccentric orbit.

The \eri\ system may also contain a circumstellar emission peak within a few
arcseconds ($< 20$~AU) of the star \citep{beichman,greaves,dent00, li03}, and probably a
second planet! Precise Doppler measurements suggest that a massive planet
($m \sin{i}=0.86\pm0.05 M_{jupiter}$) orbits \eri\ in an eccentric
orbit ($e=0.6\pm0.2$) at semimajor axis $a=3.4\pm0.1$~AU
\citep{camp88,cumm99,hatzes}. Interpretation of this data remains
controversial, partially due to the fact that \eri\ is young and 
active \citep{gray95},
despite a preliminary astrometric detection of the planet
\citep{gate00}.

The dust responsible for the emission interior to $\sim
20$~AU would eventually spiral past this
precise-Doppler planet, which could easily trap dust temporarily
in MMRs. Present images can barely resolve the disk interior to
20~AU, but future high-resolution observations could probe this
region of the debris disk and investigate the cloud structure
created by the planet. \eri's inner dust cloud may resemble the
solar zodiacal cloud perturbed by planets of the inner solar
system---though the inner \eri\ planet has an eccentric orbit.

\citet{li03} have questioned the existence of the central
submillimeter emission peak, suggesting it may be due to noise.
Initial images taken by the Spitzer Space Telescope
\citep{spitzer} of another debris disk system, Fomalhaut, reveal
mid-IR emission from warm dust occupying a central region of the
system that had appeared relatively empty in earlier submillimeter
images \citep{holland}. In addition, the excess flux detected by
IRAS at $25$~$\mu$m around \eri\ was contained in a beam only
sensitive to the inner 20~AU of the system \citep{beichman}.  
Even if the claimed central
submillimeter emission is spurious, then, it is still likely that \eri\
contains a central dust cloud; if the radial velocity planet is real, 
this cloud should reflect its presence.

The \eri\ system provides a rare opportunity to compare dust cloud
simulations to images of a nearby, roughly face-on debris disk, in
a case where some orbital parameters of the perturbing planet are
independently measurable---to build a bridge between different
dynamical methods for detecting extrasolar planets. In this paper,
we examine the inner ``exozodiacal'' component of the \eri\ dust,
near the precise-Doppler planet. We use numerical simulations of
the interaction of a planet and a dust cloud to predict what
observations might reveal at $\le 20$~AU in this system. We
consider how dust cloud imaging could test the
existence of the reported planet, and, if it exists, constrain its
properties.

\section{SIMULATION PARAMETERS}

To model the effect of the precise-Doppler planet on the dust in
\eri\ we numerically integrated the equations of motion for dust
grains \citep{liou}, using a symplectic $n$-body map
\citep{wisdom}, modified to include terms for radiation pressure
and P-R drag \citep{wilner}. We do not attempt to model the effect
of collisions on the eventual distribution of dust. In each
simulation, we integrated the orbits of 1000 dust grains for a maximum of
$5\times10^8$ years, the
approximate age of the \eri\ system.  We stop integrating a particle's orbit
if it enters the planet's Hill sphere or reaches
an orbit with eccentricity $e_{dust}>1.0$ or semimajor axis
$a_{dust}>200$~AU or $a_{dust}<0.1$~AU.
\subsection{The Planet}
\citet{hatzes} combined radial velocity data from several sources
(McDonald Observatory, Lick Observatory, the European Southern
Observatory at la Silla and the Canada-France-Hawaii Telescope)
and deduced that the planet's orbit has semimajor axis
$a=3.4\pm0.05$~AU and eccentricity $e=0.6\pm0.05$.  Radial
velocity data from Lick Observatory \citep{marcy} considered alone
yield a different eccentricity ($e=0.43$), suggesting that the
true uncertainties in the planet's orbital parameters are quite
high. We show simulations with $e=0.6$, though we tried
simulations with both values and found no significant
difference in the cloud morphology.

Our dynamical models probe the full three dimensional structure of the
dust cloud.  They require as input the inclination, $i$, and the longitude
of ascending node, $\Omega$, of the planet, orbital elements unconstrained
by radial-velocity measurements.
For the purposes of this exploration, we used $i$ and $\Omega$ from
preliminary astrometric work by \citet{gate00}.  Our models will
ultimately test the astrometric measurements.

We assume the inclination of the planet's orbit to the line of
sight is $i=46^\circ$, which \citet{gate00} determined to be the
best fit to astrometric and radial velocity data.  The
\citet{gate00} value has a large uncertainty ($\pm17^\circ$), so
it is roughly consistent with values suggested by the shape of the
outer ring of 850~$\mu$m emission \citep{greaves} and the tilt of
the stellar pole \citep{saar}.  The amplitude of the radial
velocity variations constrains the mass of the planet to $m
\sin{i}=0.86\pm0.05 M_J$ \citep{hatzes} given the mass of the star
($M_{*}=0.8 M_{\odot}$).  For $i=46^\circ$, this gives $m=1.20
M_J$.  We discuss the effects of varying planet mass and
inclination in \S~4. We adopt a longitude of ascending node 
of $\Omega=120^\circ$
\citep{gate00}, and an argument of pericenter of $\varpi=49^\circ$
\citep{hatzes}. These parameters are only weakly constrained by
data, but variations affect only the observed orientation of any
dust cloud structure.

\subsection{The Dust Grains}
Recent models \citep{sheret04, li03, dent00} attempt to use the
\eri\ spectral energy distribution (SED) and the
azimuthally-averaged radial profile from the 850~$\mu$m image to
constrain dust grain properties. However, the grains in the center of
the cloud may not resemble the grains in the outer ring; they are
likely to be smaller and more collisionally processed. 
We adopt a dust grain size of $s=15$~$\mu$m for all of
our simulations, a typical grain size for the solar zodiacal cloud. 
We discuss the effects of varying grain size in \S 4.

For spherical grains, the radiation pressure factor, $\beta$,
which determines the strength of the P-R force, is related to the
grain size by $\beta=3L_{*}/\left(16\pi cGM_{*}\rho s\right)$,
where $s$ is the grain radius in microns and $\rho$ is the dust
grain density \citep{wyat50,burn76}.  Assuming $\rho=2$~g
cm${}^{-3}$ gives $\beta = (\hbox{0.099~$\mu$m}) /s$ for grains
orbiting \eri. For 15~$\mu$m grains, this corresponds to
$\beta=0.007$.

In order to generate snapshots of the dust distribution at several
different phases of the planet, we use a procedure nearly identical 
to that used by \citet{wilner} to create face-on disk models. \citet{kuchhol} used a similar procedure to schematically illustrate
disk structures. \citet{quillen} also created disk models in this fashion, 
but restricted their simulations to two dimensions. 
We record the position of each
particle throughout the integration at regular intervals
corresponding to four particular phases on the planet's orbit,
rotating the three-dimensional positions into a coordinate system
tilted 46${}^{\circ}$ from face-on. Then we sort these positions
into four three-dimensional histograms that model the density
distribution of dust in an inertial frame at each of the four
orbital phases, assuming a steady-state source.

We calculate the emission from the density distribution using
ZODIPIC\footnote{This IDL software package is available at
http://cfa-www.harvard.edu/\char126 mkuchner} \citep{zodipic,
danc03}, assuming the cloud is optically thin, and we scale the
column density so the total 25~$\mu$m flux matches that measured
by IRAS.  As mentioned above, at this wavelength IRAS was
sensitive to emission within approximately 20~AU of \eri, the size
of our simulated images.

ZODIPIC iteratively calculates dust grain temperatures for grains
in equilibrium with stellar radiation. We model the photospheric
emission from the star by a blackbody at 5156K \citep{bell89} with
luminosity $L_{*}=0.34L_\odot$ and distance 3.218~pc. We assume a
simple modified blackbody model for dust grain emission.  At wavelengths
longer than some critical wavelength, $\lambda_0$, a dust grain's
emission coefficient, $\epsilon$, decreases roughly according to a
power law; $\epsilon=\epsilon_0\left(\lambda_0/\lambda\right)^q$. At
wavelengths shorter than $\lambda_0$, we hold $\epsilon$ constant.
We assume amorphous, moderately absorbing grains, for which
$q=1.0$ and $\lambda_0$ equals the grain radius \citep{backman}.
This choice falls between the porous grains favored by
\citet{li03} and the more solid grains favored by
\citet{sheret04}. We discuss the uncertainty in dust emission
properties in \S~4.

The initial particle inclinations $i'$ were chosen from a uniform
distribution on the interval $[0,5]^\circ$, where $i'=0^\circ$
corresponds to the plane of the planet's orbit. The initial
ascending node, argument of periapse, and mean anomaly for each
particle were selected from a uniform distribution on the interval
$[0,2\pi)$. We explored a range of initial particle
eccentricities.

\section{SIMULATION RESULTS}
Since the P-R timescale for dust around \eri\ is less than the age of
the system, none of the observed dust can be considered primordial;
its presence requires the existence of
dust-generating bodies in the system.
We present the results of two simulations with different initial
conditions, meant to model dust grains 
generated by different populations of small bodies. We choose to study 
two populations that have analogs in our solar
system, one where dust grains begin on low-eccentricity orbits, and
one where they begin on highly eccentric orbits. Later in this
section, we discuss the parameter space between these two
extremes. 

\subsection{Asteroid Dust Model}

\begin{figure}
\plotone{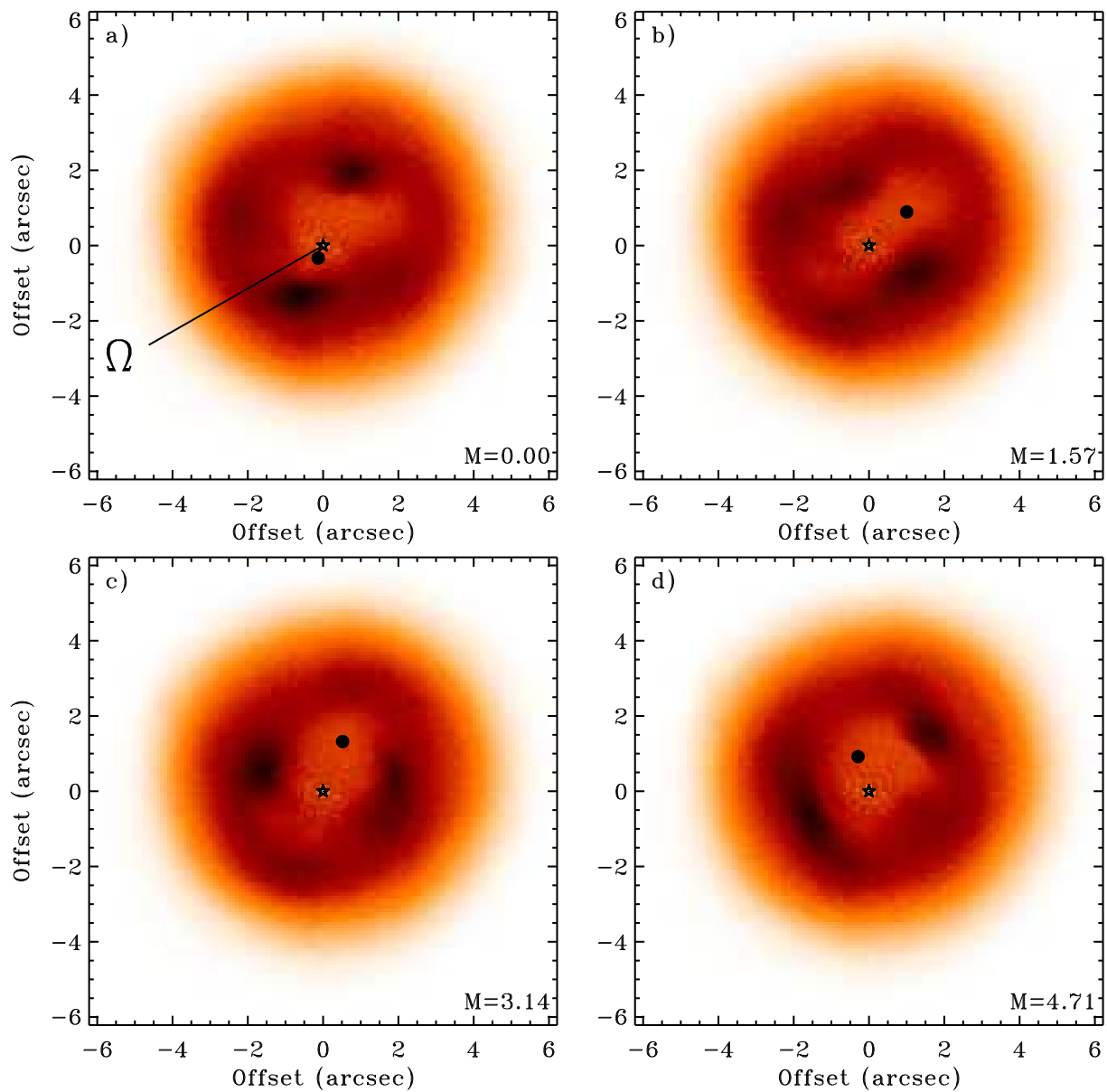} \caption{ \label{fig:snapshots}Simulated dust
distribution for source grains beginning on low eccentricity
orbits.  The snapshots show the structure at four different
orbital phases of the planet: a) $M=0$, b) $M=\pi/2$, c) $M=\pi$,
d) $M=3\pi/2$.  A filled circle represents the planet, and a star
symbol denotes the star.}
\end{figure}

For the first model, we started the grains on orbits with
semimajor axes $a=15$~AU and eccentricities chosen from a uniform
distribution on the interval $[0.05,0.15]$. This model represents
dust generated in a belt of small bodies on low-inclination orbits
like the asteroid belt.   
Figure~\ref{fig:snapshots} shows four snapshots of the model at
850~$\mu$m, when the planet's mean anomaly is $M=0$, $\pi/2$,
$\pi$ and $3\pi/2$.

The results resemble the model for the Vega disk by
\citet{wilner}. The primary observable structure the planet carves
in the dust distribution is a pair of clumps that appears to orbit
the star once for every two orbits of the planet. The clumps occupy 
circumstellar radii of roughly
5--10~AU. This zone corresponds roughly to the orbit of dust
with $e_{dust}=0.4$ in the 3:1 exterior MMR.

\begin{figure}
\plotone{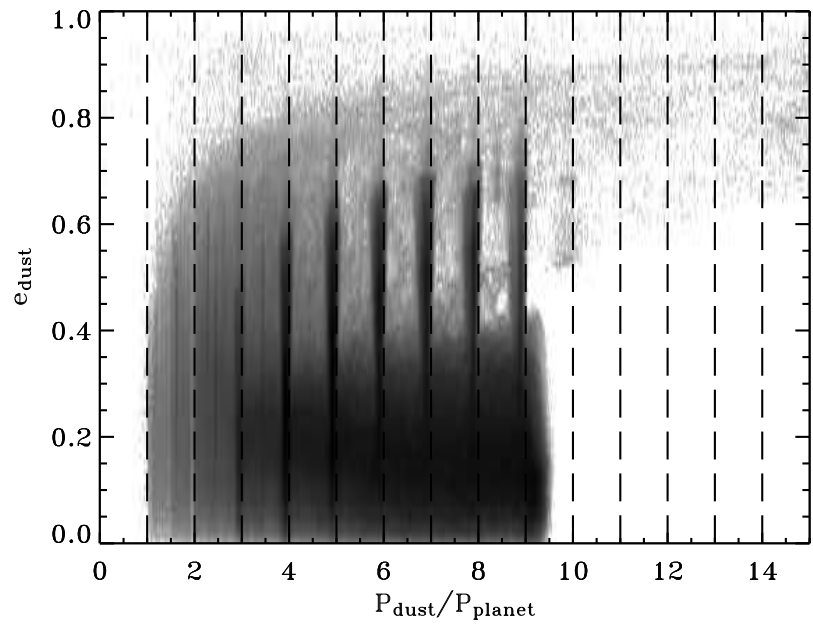} \caption{ \label{fig:pvseasteroid} Histogram of
relative dust concentration as a function of period and
eccentricity for the model shown in Figure~\ref{fig:snapshots}. Dust
concentration is displayed on a logarithmic scale.}
\end{figure}

Figure~\ref{fig:pvseasteroid} shows a histogram of the dust concentration
as a function of eccentricity and orbital period
for the particles represented in Figure~\ref{fig:snapshots}.  The
dust was released at $a=15$~AU, where the dust orbital period, $P_{dust}$,
is approximately 9.3 times the orbital period of the planet, $P_{planet}$.
In  Figure~\ref{fig:pvseasteroid}, particles trapped in MMRs
form dark vertical bars; the bars at $P_{dust}/P_{planet} = 3,4,5,\ldots,9$
show that the resonantly trapped grains in this model primarily occupy
the 3:1 through 9:1 exterior MMRs.  Fainter dark lines are also visible
at the periods of other resonances, notably the 5:2 and 7:2
exterior MMRs. Dust is captured most strongly into the 3:1 and 4:1 resonances.

\clearpage

\clearpage

\citet{khsignatures} illustrated four principal structures a
planet could create in an optically thin disk of dust released on
low-eccentricity orbits.  Our asteroidal dust model for \eri\ corresponds to
case IV in that paper, the high-mass planet on a moderately eccentric orbit.
The \citet{quillen} model of the outer ring
corresponds to case III; \eri\ could host
both a case III dust ring and a case IV dust ring.
Figure~\ref{fig:pvseasteroid} illustrates that the eccentricities
of the dust grains in the MMRs range from 0 to roughly the planet's
eccentricity (0.6) as described in \citet{khsignatures}.

Besides the dust clumps, Figure~\ref{fig:snapshots} shows a torus
of dust particles that is not centered on the star.  This torus
appears to be generated at the same time as the clumps by the
mechanism described in \citet{khsignatures}.
Figure~\ref{fig:pvseasteroid} reveals that most of the dust in
non-resonant orbits is concentrated in orbits with $e_{dust} <
0.4$. But the offset of the center of the torus ($\sim 0.5''$)
from the star shows that it must contain higher eccentricity
particles, such as those in the most distant $n:1$ MMRs.

To first order, changing the mass of the planet, the initial semimajor
axis of the dust or $\beta$ only changes which resonances are populated;
e.g., for more massive planets, more distant resonances dominate.  However, for
planets much more massive than Neptune, the dominant MMRs are always of
the form n:1.  Since MMRs of the form n:1 generally produce a
two-clump structure, the appearance of the cloud
hardly  changes whether MMRs 2:1 to 5:1 dominate or  MMRs 5:1-8:1
dominate, for example.
Consequently, the basic resonant pattern is robust to changes in the above
mentioned
parameters.

\subsection{Comet Dust}

We also considered dust released at higher
orbital eccentricities.  Such dust could be generated by small bodies
in eccentric orbits, like comets in our solar system.
The outer ring may scatter such objects inward just as the Kuiper Belt leaks
comets into the inner solar system \citep{fern80,dunc88,levi97}.

\begin{figure}
\plotone{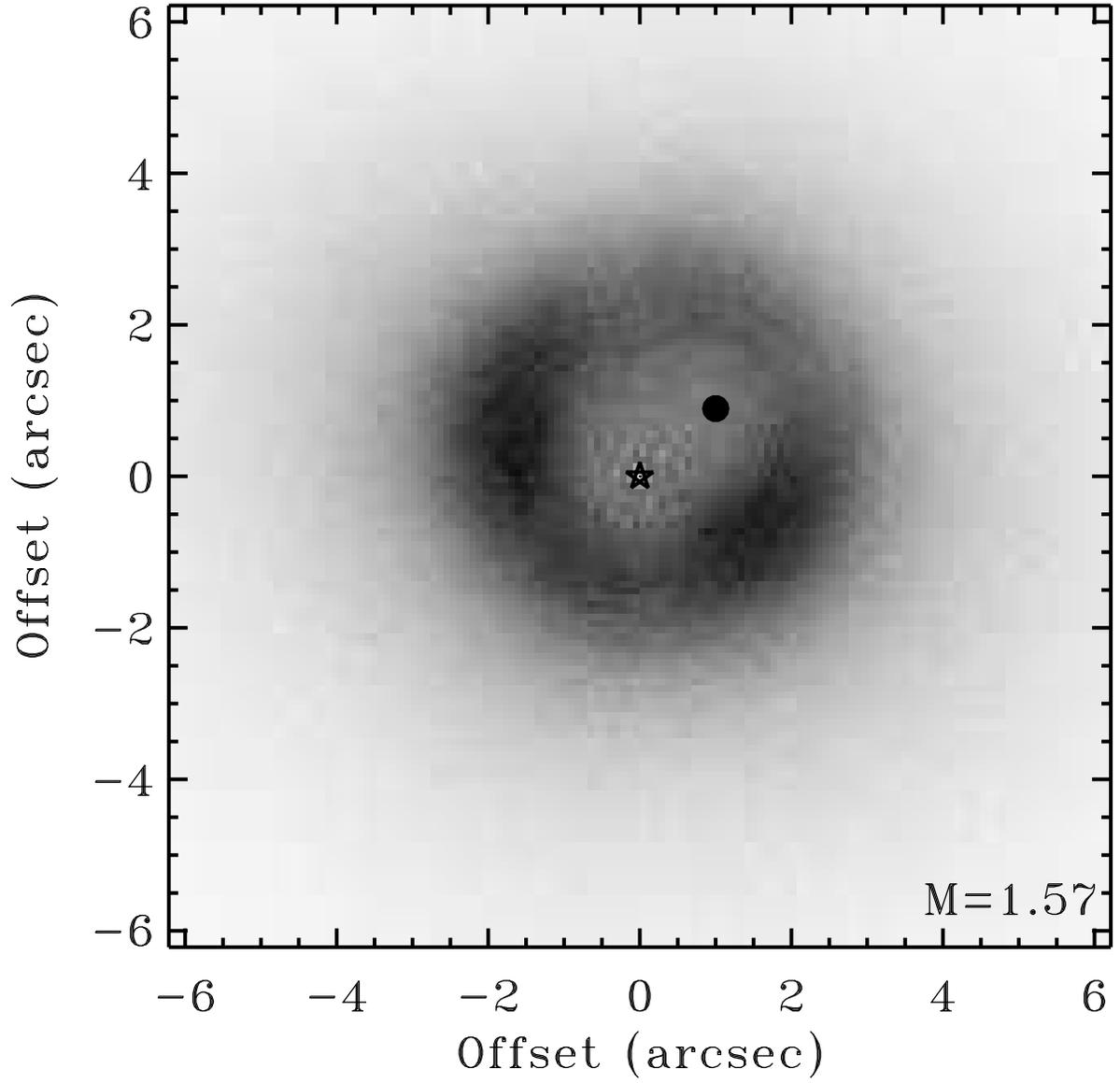} \caption{ \label{fig:cometcloud}Simulated dust
distribution for source grains beginning on high eccentricity
orbits with semimajor axes of 50~AU.  The structure does not vary
with the planet's orbital phase.}
\end{figure}

Figure~\ref{fig:cometcloud} shows an extreme example, a 1000
particle model with initial dust semimajor axes $a_{dust}=50$~AU
and eccentricities chosen from a uniform distribution on the
interval $[0.8,0.9]$. The choice of semimajor axis is motivated by
the location of the Kuiper Belt in the solar system. For such high
eccentricities, however, moderate changes in initial semimajor
axis will do little to alter a particle's steep initial approach
to the center of the system.

This model generates a substantially different cloud morphology
than the asteroid dust model described above. This comet-dust
model has a torus of emission, but no resonant dust clumps.  The
inclination of the torus produces two limb-brightened areas where
the optical depth through the torus is highest.  These
enhancements resemble the rotating clumps of
Figure~\ref{fig:snapshots}, but unlike the clumps generated by
dust in MMRs, the emission enhancements due to limb-brightening do
not vary with time.

\clearpage

\clearpage

\begin{figure}
\plotone{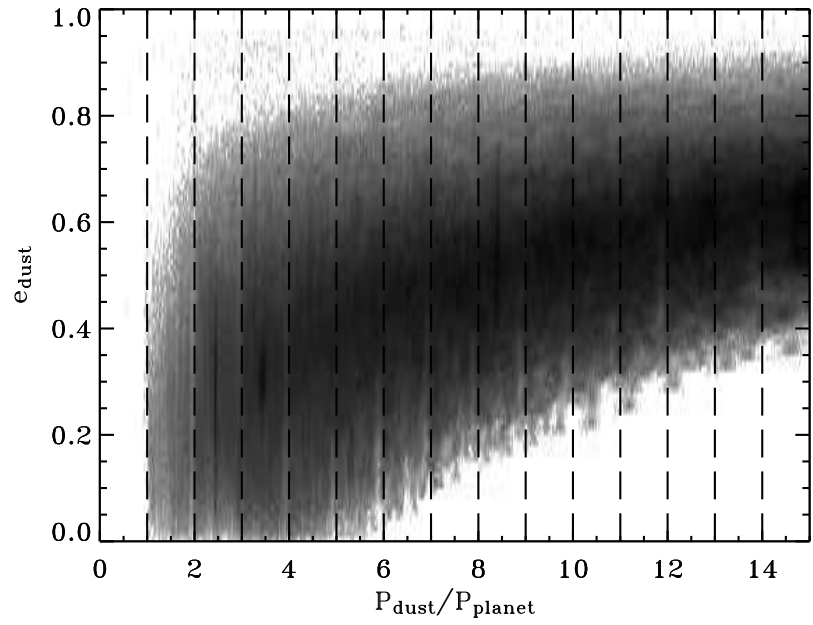} \caption{ \label{fig:pvsecomet} Histogram of
relative dust concentration as a function of period and
eccentricity for the comet dust model shown in
Figure~\ref{fig:cometcloud}.}
\end{figure}

Figure~\ref{fig:pvsecomet} shows the histogram of dust
concentration as a function of eccentricity and orbital period for
the dust in Figure~\ref{fig:cometcloud}. This histogram lacks the
dark vertical bars that appear in Figure~\ref{fig:pvseasteroid}
indicating dust trapped in the 3:1 through 9:1 MMRs. These regions
of the histogram in Figure~\ref{fig:pvsecomet} are
relatively unpopulated, suggesting that the dust released at high
orbital eccentricities avoids these n:1 MMRs.  Faint dark bars appear at the
locations of several other MMRs, but they represent only weak
enhancements over the background of particles in non-resonant orbits.

\clearpage

\clearpage

Because fewer comet-dust particles become trapped in MMRs, dust grains
in the comet-dust simulation pass through the central 6'' more
rapidly, on average, than grains in the asteroidal dust simulation described
above. The mean lifetime of a dust grain in
Figure~\ref{fig:cometcloud} is $\tau \sim 5.5\times10^6$~yrs, compared to
$\tau \sim 8\times10^6$~yrs for the dust in Figure~\ref{fig:snapshots}.
This shorter dust lifetime reduces the amount of information that goes
into each snapshot of dust density.  As a result,
Figure~\ref{fig:cometcloud} appears grainier than
Figure~\ref{fig:snapshots}.

The resonant dust clumps seen in Figure~\ref{fig:snapshots} seem to
emerge only when the dust originates on low-eccentricity
orbits.  By running a set of brief simulations of 100 particles
each, we traced the initial dust grain eccentricity
at which the resonant clumps begin to vanish.
Dust grains released with eccentricities as high as 0.4
continue to be trapped into the n:1 MMRs, but the clumps become fainter and
fainter as the initial $e_{dust}$ is increased.  At initial
dust orbital eccentricities greater than $e_{dust} \approx 0.6$, our
simulations no longer show any sign of trapping in the n:1 MMRs.

This critical eccentricity is the eccentricity of the planet's orbit.
The disappearance of clumps may simply reflect when the orbits of particles in
the key MMRs begin to intersect the orbits of the planet.
Particles in the 3:1 exterior mean-motion resonance
would become planet crossing at some longitudes of pericenter at
an eccentricity of $e_{dust} \approx 0.2$.  At $e_{dust} \approx 0.8$,
all particles at the nominal semimajor axis of this
resonance must cross the planet's orbit.

\section{DISCUSSION}

To match the observed total emission from dust in the
central $6''$ of the \eri\ cloud, our simulations require a mass
of $0.1$ lunar masses ($\sim7\times10^{24} g$) of $15$~$\mu$m dust
grains. The mean line-of-sight optical depth for both asteroid and comet
dust models is $5\times10^{-3}$. Both the resonant blobs of the
asteroid model and the limb-brightened regions of the comet model
have line-of-sight optical depths of approximately $0.02$.

A better calculation of the 
dust mass in the system's center depends
on a realistic model for dust grain size distribution and physical
characteristics (porosity, etc). Models similar to those in
\citet{li03} and \citet{sheret04} might give more accurate mass
estimations, but due to the lack of observational constraints, none yet
exist for the central zodiacal cloud of \eri.  Regardless of the
physical model used to calculate dust mass, most of the mass in the
debris disk will reside in the source population of small bodies.

\begin{figure}
  \plotone{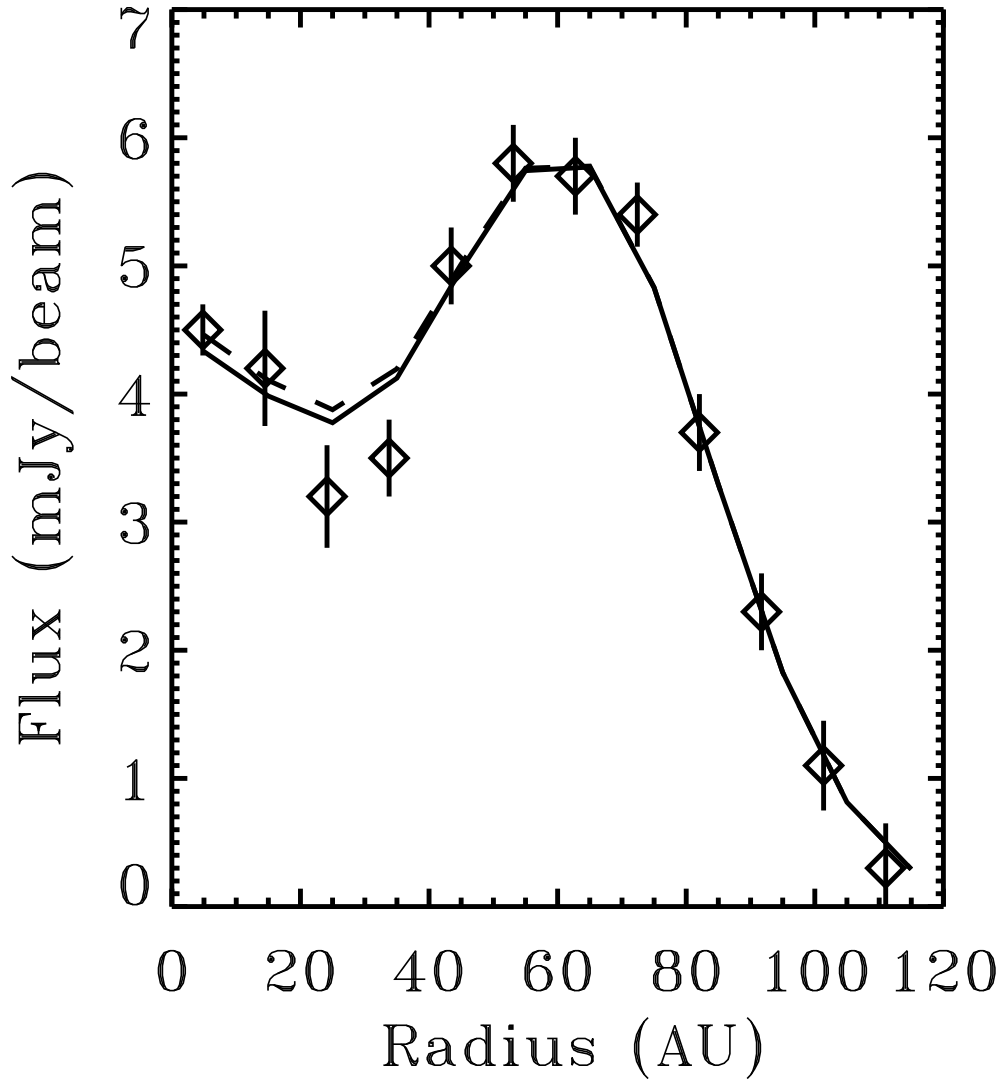} \caption{\label{fig:radprofile}The azimuthally
  averaged radial profile of the \eri\ dust disk at 850~$\mu$m. 
  Open diamonds with error bars are from \citet{greaves}. Solid line
  is the radial profile of our asteroid dust model, convolved with a
  Gaussian of 15'' FWHM. The dashed line is the same, for our comet
  model. For clarity, we include the radial
  profile of a simple symmetric ring centered at 72 AU, 
  consisting of grains with the
  same properties as our inner dust models. Both asteroid and comet
  model are consistent with the inner radial profile ($<20$~AU).}
\end{figure}

Our simulated dust clouds are nevertheless consistent with current
observations of the inner dust cloud, for both the asteroid and comet
models. Based on the dust grain 
physical properties and emission characteristics discussed in \S~2.2, 
we generate 850~$\mu$m, azimuthally averaged, radial profiles of our 
simulated dust clouds.
These are convolved with a Gaussian of full-width half-maximum
(FWHM) 15'' to match the resolution of the
\citet{greaves} image. Figure~\ref{fig:radprofile} presents these
radial profiles, plotted against the \citet{greaves} radial
profile. Both models are consistent with the observed profile.
In addition, the total 850~$\mu$m flux within 20~AU implies a  
total flux at 25~$\mu$m that is consistent with the 25~$\mu$m IRAS 
excess \citep{beichman}.

\clearpage

\clearpage

When the planet mass is varied by $\pm 0.4 M_J$, a range
consistent with current uncertainties in inclination and
$m\sin{i}$, as well as the uncertainty in the mass determination
of \citet{gate00}, we find no significant difference in the cloud
morphology. 
Aside from its effect on planet mass, varying the inclination of
the planet's orbit would affect only the projected separation
between any dust blobs (or ring) and the central star. However, only in the
extreme case of a nearly edge-on planetary orbit would the
observed dust morphology be unrecognizable; since all observations
of \eri\ indicate a roughly face-on orientation, we do not believe
that inclination effects are a serious concern.

\subsection{Shortcomings of the Models}

A weakness of our simulations is that
we cannot properly account for collisions between grains.
The time scale for dust grain collisions within the inner ring of
dust ($\tau_\perp=1.0\times 10^{-4}$) is long compared to collision
time scales in the debris disks
around Vega, $\beta$ Pictoris and Fomalhaut \citep{dent00}, but
still less than $10^5$ years for grains within 20~AU of the star,
shorter than the P-R time scale ($10^6$--$10^7$ years).
In the absence of larger bodies, collisions tend to limit the
maximum dust density in any given area by destroying grains.
On the other hand, grains colliding with
larger bodies can generate dust.  The short collisional lifetime
of the grains suggests that the dust may be
generated by bodies closer to the star
than the 15~AU asteroid belt we simulated, bodies
that are perhaps themselves in resonant orbits with the planet.

Our simulations were performed with dust grains having a uniform value
of $\beta$, a consequence of adopting a uniform dust grain size. In
general, increasing
$\beta$, or equivalently decreasing the grain size, increases the
sharpness of features in our dust images. The
reverse is true when $\beta$ is decreased (grain size increased). A
more physically realistic simulation would include a range of dust
grain sizes; because the sharpness of features depends on $\beta$, we
expect actual observations to reveal features that
are somewhat less sharp than the ones presented here.

The precise-Doppler planet would interact secularly with the
planet predicted by \citet{quillen} on a time scale of
$10^8$--$10^{9}$ years. The structure of the outer ring, with its
longer P-R time, may reflect this interaction, though the secular
perturbations of the outer planet on the inner planet and inner
dust ring probably have little consequence.   However, the system could
contain other unknown planets, which could add significant
perturbations.

\subsection{Future Observations}

\begin{figure}
\plotone{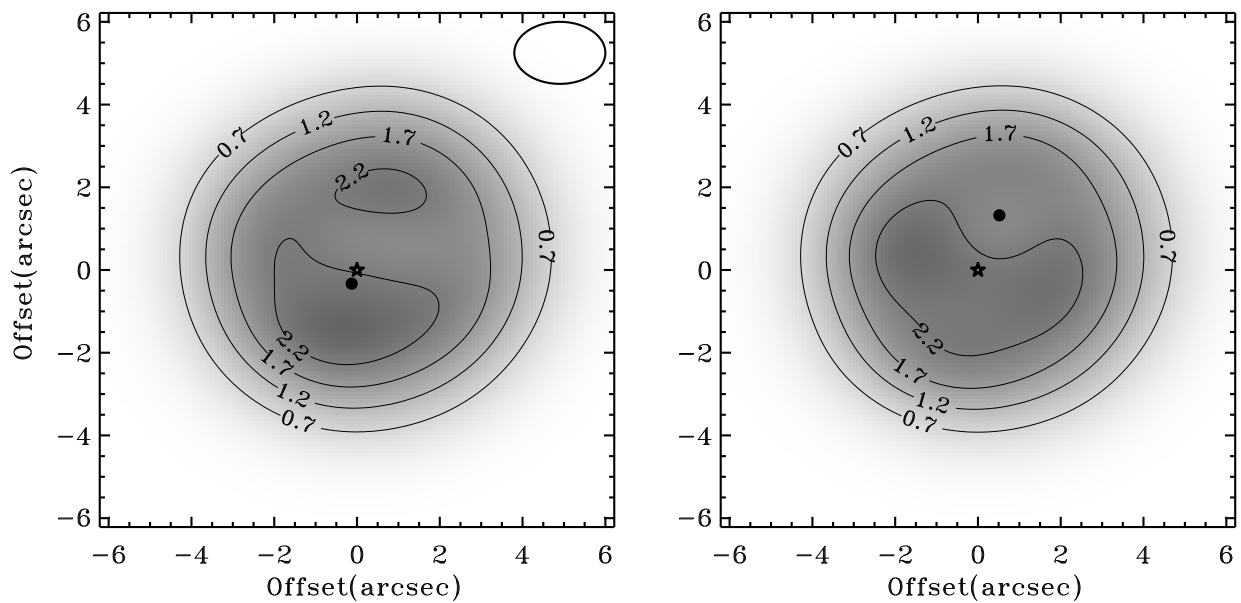} \caption{\label{fig:simobs} Simulated
observations of the asteroid dust model at two different phases of
the planet's orbit, as seen by the IRAM PdBI at 1.3mm.  Contours
indicate the surface brightness, in mJy beam${}^{-1}$. The
locations of the blobs reflect the planet's argument of pericenter
($49^\circ$).}
\end{figure}

\begin{figure}
\plotone{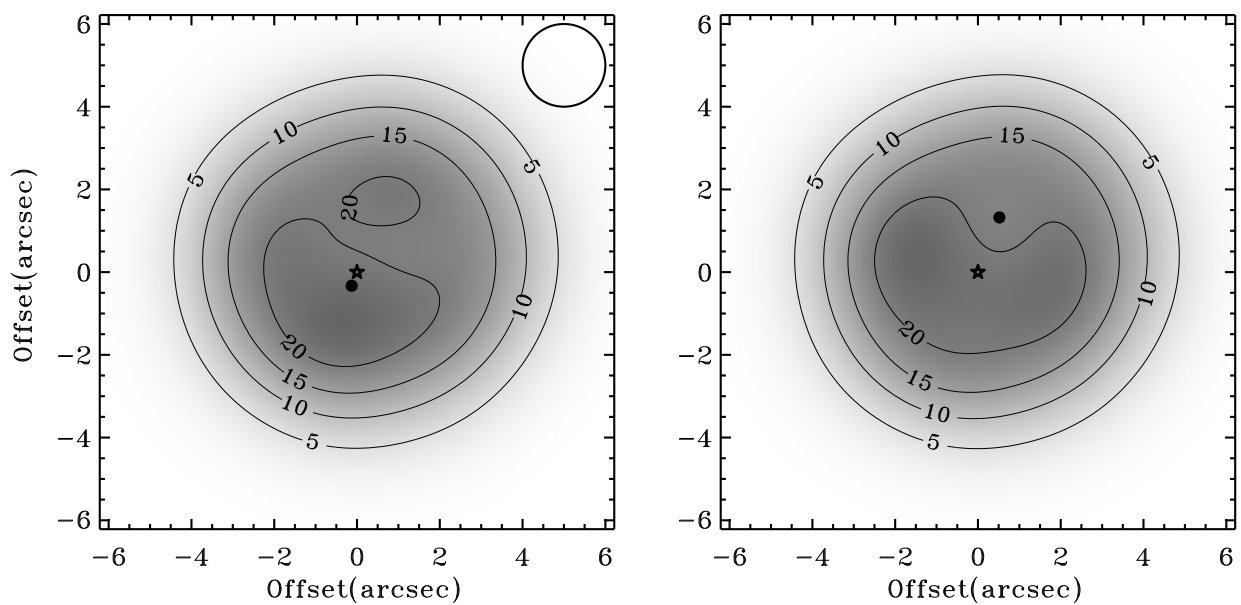} \caption{\label{fig:simobs2} Simulated
observations of the asteroid dust model as seen by the SMA at
450$\mu$m. Contours indicate the surface brightness, in mJy
beam${}^{-1}$.}
\end{figure}

Observations of the inner regions of \eri\ with future telescopes might
allow us to not only confirm the existence of the planet, but also
to infer the existence of an asteroid belt or some other source of
dust located within the outer dust ring.
To assess the detectability of the expected dust morphology, we
convolved our simulated images with Gaussians representing the
beams of several current and planned telescopes.
Figure~\ref{fig:simobs} presents convolved images of the dust
structure at two different phases of the planet, as the IRAM
Plateau de Bure Interferometer (PdBI) might see at 1.3mm. We simulate
an elliptical beam for PdBI, $1.5'' \times 2.2''$ FWHM, appropriate for a
source at the declination of \eri, observed by PdBI in the ``CD''
configuration.
Figure~\ref{fig:simobs2} shows similar images for the Submillimeter
Array (SMA) at 450~$\mu$m. In each image, we remove 90\% of the starlight
before convolution.

\clearpage

\clearpage

Observations with PdBI at 1.3mm may marginally resolve the dust
blobs. The expected emission at 1.3mm, based on the IRAS photometry, 
is very close to
the sensitivity limit of PdBI ($0.7$ mJy beam${}^{-1}$). At this
level, the probability of detection depends critically on the
physical model used to predict dust emission at 1.3mm from the
observed $25\mu$m flux. Deviations from our modified blackbody
emission model could reduce the expected flux to a level below the
detection threshold, or enhance it to provide a higher-confidence
detection.

Observations with the SMA at 450$\mu$m should provide higher
angular resolution, though the expected flux levels remain close
to the detection limit ($\sim 8$~mJy beam${}^{-1}$). We expect the
SMA to easily separate the dust emission from the stellar
emission, though it may still be difficult to simultaneously
achieve the sensitivity and resolution needed to resolve
individual dust clumps.  In general, an instrument must achieve a
resolution of better than $\sim 3$ arcseconds in order resolve the
dust blobs.

Though Spitzer is scheduled make observations of the \eri\ circumstellar
disk, the proximity of
our dust blobs to the central star will make them difficult to
separate from the stellar glare. At wavebands where Spitzer may achieve
the necessary 3 arcsecond resolution, the photospheric emission
from the star will most likely obscure any faint emission from
the dust.

\begin{figure}
\plotone{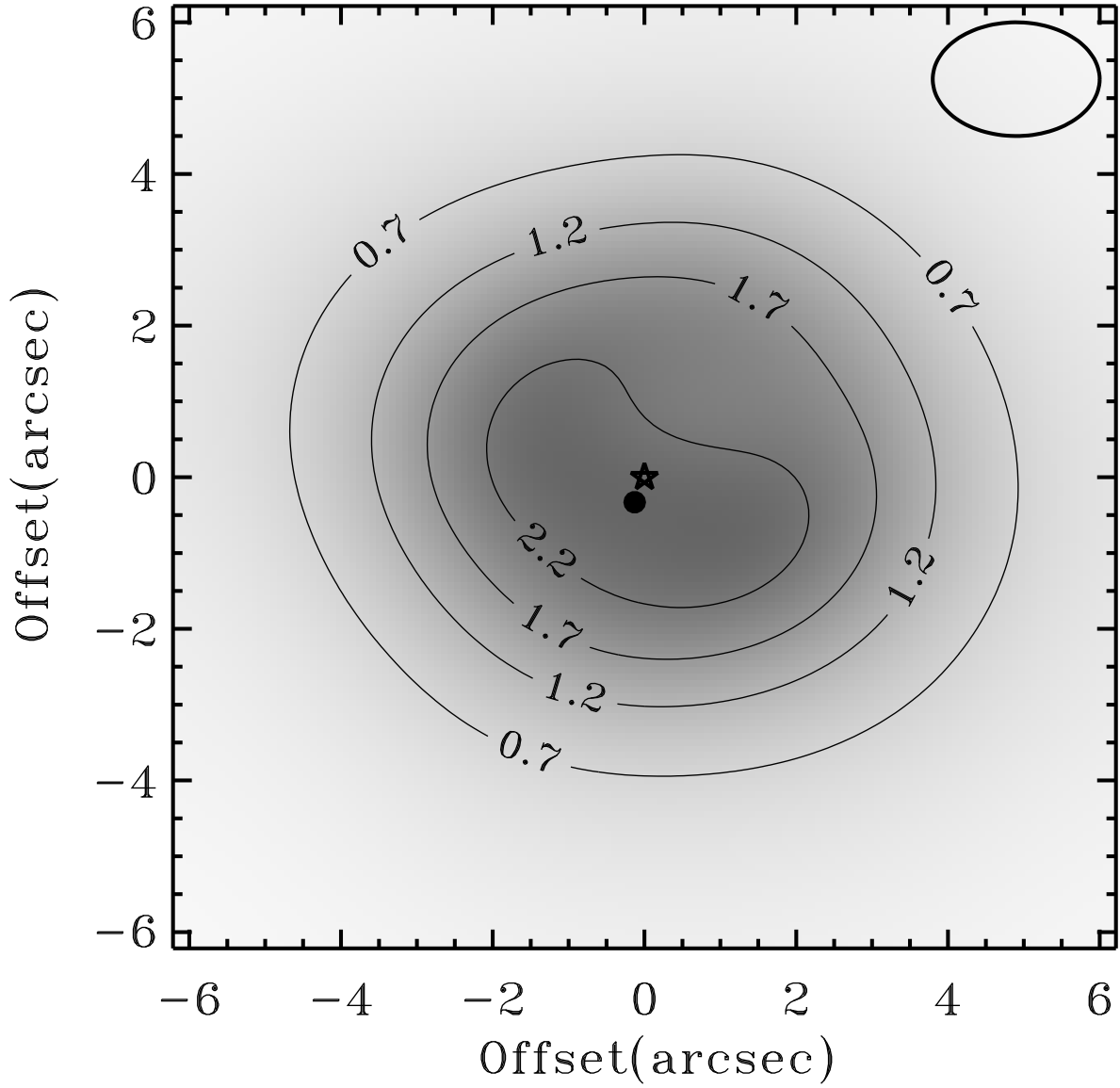} \caption{\label{fig:cometsim1} Simulated
observations of the comet dust model by IRAM PdBI at 1.3mm.
Contours indicate the surface brightness, in mJy beam${}^{-1}$.}
\end{figure}

\begin{figure}
\plotone{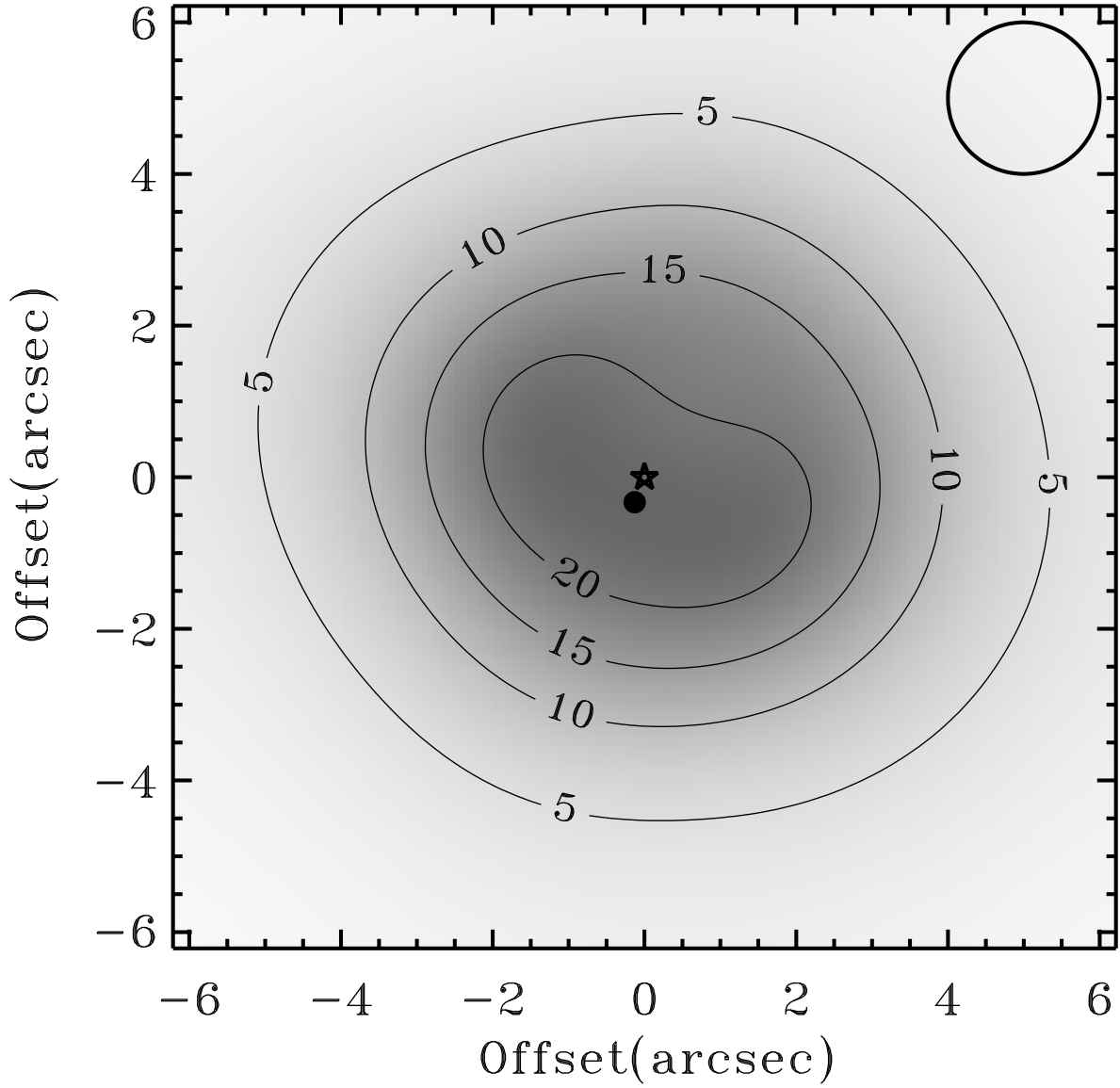} \caption{\label{fig:cometsim2} Simulated
observations of the comet dust model by SMA at 450 $\mu$m.
Contours indicate the surface brightness, in mJy beam${}^{-1}$.}
\end{figure}

Figures~\ref{fig:cometsim1}~and~\ref{fig:cometsim2}
show simulated images of our comet dust model, as seen by the PdBI and
the SMA, respectively.  These simulated images show that
a single detection of dust blobs in \eri\ would be ambiguous. Multiple
observations spanning a few years are required to test whether or not
the dust blobs move.  Detecting revolving
clumps would support the asteroid dust model, confirming the existence
of the planet and constraining its ascending node, inclination, and longitude of pericenter.

\clearpage

\clearpage

Yet even if the revolving clumps are not found, we may still be
able to use the cloud morphology to confirm the existence of the
planet. As Figure~\ref{fig:cometcloud} shows, the torus of dust
that develops when we have eccentric particles has an inner void
that is approximately circular, but the circle is not centered on
the star.  Detecting that the inner cutoff of the dust cloud
is a circle not centered on the star would also be a sign of planetary
perturbations, and it would constrain the planet's ascending node, inclination, and
longitude of pericenter.

The Atacama Large Millimeter Array (ALMA) should provide greatly
enhanced sensitivity, along with spatial
resolution equal to that of the simulated images in
Figures~\ref{fig:snapshots}~and~\ref{fig:cometcloud}.
This array should easily resolve any dust structure created by
the planet in the central 20~AU of the \eri\ system, and provide
the dynamical information described above with a single observation.
The proposed space missions SPECS \citep{leis99}, Eclipse \citep{trau03},
and TPF\footnote{See http://planetquest.jpl.nasa.gov}
should also be able to make detailed maps of the central dust cloud.

\section{CONCLUSIONS}

Our simulations suggest that if the inner ring ($<20$~AU) in the \eri\
debris disk is comprised of dust released on low-eccentricity
orbits $(e_{dust} \lesssim 0.4)$ exterior to the
precise-Doppler planet orbiting \eri,
future submillimeter images of this system should detect an off-center
limb-brightened ring and a pair of dust clumps that appear to
orbit the star once every fourteen years.  If the dust
is released on highly eccentric orbits $( e_{dust} \gtrsim 0.4)$, we
would not see rotating dust clumps, but rather an inclined
torus of dust with a circular inner void off-center from the star.
High-resolution
observations of the \eri\ dust complex with the IRAM PdBI, the
SMA, or ALMA could confirm the existence of the planet reported
by \citet{hatzes}.  Our
simulations are not the last word on the
structure of this cloud; we do not take into account the mutual
collisions of dust grains.  However, finding one of the generic structures
we predict would constrain the unknown orbital
elements of the planet and lend confidence to
the practice of interpreting dust cloud patterns as resonant
signatures of extrasolar planets.

\acknowledgements

This work was performed in part under contract with the Jet Propulsion
Laboratory (JPL) through the Michelson Fellowship program funded by
NASA as an element of the Planet Finder Program.  JPL is managed for
NASA by the California Institute of Technology.

\end{document}